\documentclass[12pt]{article}%
\usepackage{amsmath,latexsym}
\usepackage{graphicx}
\usepackage{amsmath}
\usepackage{amsfonts}
\usepackage{amssymb}%
\begin{document}

\title{{On wormholes in spacetimes of embedding class one}}
   \author{
Peter K. F. Kuhfittig*\\  \footnote{kuhfitti@msoe.edu}
 \small Department of Mathematics, Milwaukee School of
Engineering,\\
\small Milwaukee, Wisconsin 53202-3109, USA}

\date{}
 \maketitle

\begin{abstract}\noindent
An $n$-dimensional Riemannian space is said
to be of embedding class $m$ if $n+m$ is the \
lowest dimension of the flat space in which
the given space can be embedded.  A spherically
symmetric spacetime of class two can be reduced
to class one by a suitable transformation of
coordinates.  Applied to wormholes, given a
well-defined shape function $b=b(r)$, the
resulting wormhole has an event horizon and is
therefore nontraversable.  On a macroscopic
scale, $b(r)$ can be replaced by $m(r)$, the
effective mass of a spherical star of radius
$r$ with $m(0)=0$, to yield a valid solution.
Spacetimes of embedding class one have been
used successfully for modeling compact
stellar objects.  On a microscopic scale,
one can invoke noncommutative geometry to
obtain a charged nontraversable wormhole,
i.e., an Einstein-Rosen bridge, and hence
a model for a charged particle.   

\end{abstract}

\section{Introduction}\label{E:introduction}

Wormholes are handles or tunnels in spacetime
connecting widely separated regions of our
Universe or different universes altogether.
Morris and Thorne \cite{MT88} proposed the
following static and spherically symmetric
line element for a wormhole spacetime:
\begin{equation}\label{E:line1}
ds^{2}=-e^{\nu(r)}dt^{2}+\frac{dr^2}{1-b(r)/r}
+r^{2}(d\theta^{2}+\text{sin}^{2}\theta\,
d\phi^{2}),
\end{equation}
using units in which $c=G=1$.  Here $\nu=\nu(r)$
is called the \emph{redshift function}, which
must be everywhere finite to avoid the
appearance of an event horizon.  The function
$b=b(r)$ is called the \emph{shape function}
since it determines the spatial shape of the
wormhole when viewed, for example, in an
embedding diagram \cite{MT88}.  (The
embedding diagram will play a critical role
in Sec. \ref{S:complete}.)  The spherical
surface $r=r_0$ is the radius of the
\emph{throat} of the wormhole.  The shape
function must satisfy the following
conditions: $b(r_0)=r_0$, $b(r)<r$ for
$r>r_0$, and $b'(r_0)\le 1$, called the
\emph{flare-out condition}.  This condition
can only be met by violating the null energy
condition (NEC)
\begin{equation}
  T_{\alpha\beta}k^{\alpha}k^{\beta}\ge 0
\end{equation}\emph{}
for all null vectors $k^{\alpha}$, where
$T_{\alpha\beta}$ is the energy-momentum
tensor.  Matter that violates the NEC is
called ``exotic" in Ref. \cite{MT88}.  In
particular, for the outgoing null vector
$(1,1,0,0)$, the violation has the form
\begin{equation}
   T_{\alpha\beta}k^{\alpha}k^{\beta}=
   \rho +p_r<0.
\end{equation}
Here $T^t_{\phantom{tt}t}=-\rho$ is the energy
density, $T^r_{\phantom{rr}r}= p_r$ is the
radial pressure, and
$T^\theta_{\phantom{\theta\theta}\theta}=
T^\phi_{\phantom{\phi\phi}\phi}=p_t$ is
the lateral pressure.  A final requirement
is asymptotic flatness: $\text{lim}
_{r\rightarrow\infty}\nu(r)=0$ and
$\text{lim}_{r\rightarrow\infty}b(r)/r=0$.

Much of our discussion is based on the
assumption that our spacetime is of
embedding class one.  So we need to recall
that an $n$-dimensional Riemannian space is
said to be of embedding class $m$ if $n+m$
is the lowest dimension of the flat space
in which the given space can be embedded
\cite{MG17, MM17, MRG17, sM17, sM16, sM19}.
We also need to recall that the exterior
Schwarzschild solution is a Riemannian
space of embedding class two.

We continue our discussion with the
following static and spherically symmetric
line element from Ref. \cite{MG17}, but using
the signature from line element (\ref{E:line1}):
\begin{equation}\label{E:line2}
ds^{2}=-e^{\nu(r)}dt^{2}+e^{\lambda(r)}dr^2
+r^{2}(d\theta^{2}+\text{sin}^{2}\theta\,
d\phi^{2}).
\end{equation}
It is shown in Ref. \cite{MG17} that this
metric of class two can be reduced to a
metric of class one; the spacetime is
thereby embedded in a five-dimensional flat
spacetime.  This reduction can
be accomplished by means of the following
coordinate transformation:
$z^1=r\,\text{sin}\,\theta\,
\text{cos}\,\phi$, $z^2=r\,\text{sin}\,\theta\,
\text{sin}\,\phi$, $z^3=r\,\text{cos}\,\theta$,
$z^4=\sqrt{K}\,e^{\nu/2}\,\text{cosh}
\frac{t}{\sqrt{K}}$, and $z^5=
\sqrt{K}\,e^{\nu/2}\,\text{sinh}
\frac{t}{\sqrt{K}}$.  The result is
\begin{equation}\label{E:line3}
ds^{2}=-e^{\nu}dt^{2}
 +\left[1+\frac{1}{4}Ke^{\nu}(\nu')^2\right]dr^2
+r^{2}(d\theta^{2}+\text{sin}^{2}\theta\,
d\phi^{2}),
\end{equation}
Metric (\ref{E:line3}) is equivalent to
metric (\ref{E:line2}) if
\begin{equation}\label{E:key}
   e^{\lambda}=1+\frac{1}{4}Ke^{\nu}(\nu')^2,
\end{equation}
where $K>0$ is a free parameter.
 Eq. (\ref{E:key}) can also be obtained from
the Karmarkar condition \cite{kK48}
\begin{equation*}
  R_{1414}=
  \frac{R_{1212}R_{3434}+R_{1224}R_{1334}}
  {R_{2323}},\quad R_{2323}\neq 0,
\end{equation*}
which is equivalent to the above reduction.
In fact, Eq. (\ref{E:key}) is a solution
to the differential equation (readily solved
by separation of variables)
\begin{equation*}
   \frac{\nu'\lambda'}{1-e^{\lambda}}=
   \nu'\lambda'-2\nu''-(\nu')^2,
\end{equation*}
so that $K$ is actually an integration
constant \cite{MM17, SF20}.
%END OF SECTION

\section{Seeking a complete wormhole solution}
     \label{S:complete}
The strategy adopted by Morris and Thorne
in Ref. \cite{MT88} was to satisfy the
geometric requirements for a traversable
wormhole by specifying $b(r)$ and $\nu(r)$
and then either manufacture or search the
Universe for matter or fields that can
produce the desired energy-momentum
tensor.  In this section, we will consider
the case where $b=b(r)$ is a legitimate
shape function, which may actually be known
for physical reasons, such as a
noncommutative-geometry background.  (This
possibility will be explored further in
Sec. \ref{S:microscopic}.)  Another
possibility is to start with a constant
energy density, as in Ref. \cite{sS05}.
Returning to Eq. (\ref{E:key}), let us
rewrite the equation as
\[
   \frac{1}{\sqrt{K}}\sqrt{e^{\lambda}
   -1}=e^{\frac{1}{2}\nu}\frac{\nu'}{2}.
\]
Integrating, we obtain from $e^{\lambda(r)}
=[1-b(r)/r]^{-1}$,
\begin{equation}\label{E:EQ1}
   \sqrt{K}e^{\frac{1}{2}\nu(r)}=
   \int\sqrt{e^{\lambda(r)}-1}\,\,dr=
   \int\left(\frac{r}{b(r)}-1\right)
      ^{-\frac{1}{2}}dr
\end{equation}
and
\begin{equation}\label{E:EQ2}
   e^{\nu(r)}=\frac{1}{K}
   \left(\int\left(\frac{r}{b(r)}-1
   \right)^{-\frac{1}{2}}dr\right)^2.
\end{equation}
This integral exists as long as $b(r)$ is
a continuous function since
\begin{multline*}
  \left(\frac{r}{b(r)}-1\right)^{-\frac{1}{2}}
  =\left(\frac{r}{b(r)}\right)^{-\frac{1}{2}}
  +\left(-\frac{1}{2}\right)\left(\frac{r}
  {b(r)}\right)^{-\frac{3}{2}}(-1)\\
  +\frac{\left(-\frac{1}{2}\right)
     \left(-\frac{3}{2}\right)}{2!}
     \left(\frac{r}{b(r)}\right)
     ^{-\frac{5}{2}}(-1)^2+
        \cdot \cdot \cdot.
\end{multline*}
This series converges for $r>r_0$,
resulting in a removable discontinuity at
$r=r_0$.  Unfortunately, the integral in
Eq. (\ref{E:EQ1}) is nothing more than the
profile curve $z(r)$ in the standard
embedding diagram in Ref. \cite{MT88}:
$z(r)$ is rotated about the $z$-axis and
in the resulting tunnel-like figure, the
circle $r=r_0$ lies in the plane $z=0$.
We conclude that there is an event
horizon for any shape function; so we do
not get a traversable wormhole.

An interesting special case is provided
by $b(r)=2M$, a zero-density wormhole:
\begin{equation}\label{E:EQ3}
   e^{\frac{1}{2}\nu(r)}=\frac{1}{
   \sqrt{K}}\int^r_{2M}\sqrt{\frac{1}
   {\frac{r'}{2M}-1}}\,dr'=\frac{4M}
   {\sqrt{K}}\sqrt{\frac{r}{2M}-1}.
\end{equation}
The wormhole spacetime is not
asymptotically flat and will have to
be cut off at some $r=a$ and joined to
an external Schwarzschild spacetime.
In other words,
\begin{equation*}
   e^{\nu(a)}=\frac{16M^2}{K}\left(
   \frac{a}{2M}-1\right)=1-\frac{2M}{a}.
\end{equation*}
Since $K$ is a free parameter, we can let
\[
  K=\frac{16M^2\left(\frac{a}{2M}-1\right)}
 {1-\frac{2M}{a}}.
\]
The result is
\begin{equation}
   e^{\nu(r)}=\frac{1-\frac{2M}{a}}
   {\frac{a}{2M}-1}\left(
   \frac{r}{2M}-1\right),
\end{equation}
confirming the existence of an event
horizon at $r=2M$; also, observe that
$e^{\nu(a)}= 1-\frac{2M}{a}$.  (A similar
junction condition would be needed for
any shape function.)

Returning to Eq. (\ref{E:key}), we have
seen that whenever we assume a wormhole \
structure with a well-defined shape
function, we cannot avoid an event
horizon.  On the other hand, if we reverse
our point of view by assuming that $\nu(r)$
is finite (i.e., no event horizon), then
Eq. (\ref{E:key}) yields
\begin{equation}
   b(r)=r\left(1-\frac{1}
   {1+\frac{1}{4}Ke^{\nu(r)}[\nu'(r)]^2}
      \right);
\end{equation}
but in order to satisfy the condition
$b(r_0)=r_0$,
the fraction inside the parentheses must
vanish.  Since $e^{\nu(r_0)}$ is finite,
we must have $\nu'(r_0+)=\pm\infty$.  This
is entirely possible, as can be seen from
the choice $\nu(r)= \pm 2\sqrt{r-r_0}$.
However, the resulting wormhole behaves
much like a Schwarzschild black hole:
if we denote $1-\frac{2M}{r}$ in the
Schwarzschild line element by
$e^{{\nu_1(r)}}$, then
$\nu_1(r)=\text{ln}\,(1-\frac{2M}{r})$
and
\begin{equation*}
   \text{lim}_{r\rightarrow 2M+}\nu'_1(r)=
   \text{lim}_{r\rightarrow 2M+}
   {\frac{\frac{2M}{r^2}}{1-\frac{2M}{r}}}
   =+\infty.
\end{equation*}

We conclude that a four-dimensional
Riemannian space of embedding class one
does not allow a traversable wormhole.
For traversability, some additional
assumptions would be needed.  For example,
Ref. \cite {pK18} assumes conformal
symmetry, while Ref. \cite{pK19} uses
a modified shape function.
%END OF SECTION

\section{A stellar model}
Since the embedding theory has failed to
produce a macroscopic traversable
wormhole, let us consider instead the
stellar model \cite{MTW}
\begin{equation}\label{E:line4}
ds^{2}=-e^{\nu(r)}dt^{2}+\frac{dr^2}{1-m(r)/r}
+r^{2}(d\theta^{2}+\text{sin}^{2}\theta\,
d\phi^{2});
\end{equation}
here $m(r)$ is the effective mass inside a
spherical star of radius $r$ with $m(0)=0$.
For this model, we can return to Eq.
(\ref{E:EQ2}) to deduce that
\begin{equation}\label{E:EQ4}
   e^{\nu(r)}=\frac{1}{K}\left(
   \int^r_0\sqrt{\frac{1}
   {\frac{r'}{m(r')}-1}}\,dr'\right)^2.
\end{equation}
If the star has radius $r=R$, then the
free parameter $K$ once again allows a
junction to an external Schwarzschild
spacetime:
\begin{equation}
    e^{\nu(R)}=\frac{1}{K}\left(
   \int^R_0\sqrt{\frac{1}
   {\frac{r}{m(r)}-1}}\,dr\right)^2
   =1-\frac{2M}{R},
\end{equation}
where $M=\frac{1}{2}m(R)$.  So
\begin{equation}
  K=\frac{\left(\int^R_0
  \sqrt{\frac{1}{\frac{r}{m(r)}-1}}\,dr
  \right)^2}{1-\frac{m(R)}{R}}.
\end{equation}
For the potential $\nu$, we have
\begin{equation}
   \nu=\text{ln}\left(1-\frac{m(R)}{R}
   \right)\approx -\frac{m(R)}{R}.
\end{equation}
So for large $r$, $\nu=-\frac{m}{r}$,
the Newtonian limit.  Spacetimes of
embedding class one  have proved to
be very effective for modeling compact
stellar objects such as neutron
stars and pulsars \cite {MG17}.
%END OF SECTION

\section{Microscopic wormholes}
   \label{S:microscopic}
A convenient way to study microscopic
wormholes is by means of noncommutative
geometry \cite{GL09}.  An important
outcome of string theory is the
realization that coordinates may
become noncommutative operators on a
$D$-brane \cite{eW96, SW99}.
Noncommutativity replaces point-like
objects by smeared objects \cite{SS03,
NSS06, NS10} with the aim of eliminating
the divergences that normally appear
in general relativity.  As a consequence,
spacetime can be encoded in the commutator
$[\textbf{x}^{\mu},\textbf{x}^{\nu}]
=i\theta^{\mu\nu}$, where $\theta^{\mu\nu}$
is an antisymmetric matrix that determines the
fundamental cell discretization of spacetime
in the same way that Planck's constant
discretizes phase space \cite{NSS06}.  An
effective way to model the smearing is
to assume that the energy density of the
static and spherically symmetric and
particle-like gravitational source is
\begin{equation}\label{E:rho}
  \rho(r)=\frac{\mu\sqrt{\beta}}
     {\pi^2(r^2+\beta)^2},
\end{equation}
which can be interpreted to mean that
the gravitational source causes the mass
$\mu$ of a particle to be diffused
throughout the region of linear dimension
$\sqrt{\beta}$ due to the uncertainty;
so $\sqrt{\beta}$ has units of length.

Next, from Eq. (\ref{E:rho}) and the Einstein
field equation
\begin{equation}\label{E:Einstein1}
  \rho(r)=\frac{b'}{8\pi r^2},
\end{equation}
 we obtain the shape
function
\begin{multline}\label{E:shape}
   b(r)=\int^r_{r_0}8\pi(r')^2\rho(r')dr'\\
   =\frac{4m\sqrt{\beta}}{\pi}
  \left(\frac{1}{\sqrt{\beta}}\text{tan}^{-1}
  \frac{r}{\sqrt{\beta}}-\frac{r}{r^2+\beta}-
  \frac{1}{\sqrt{\beta}}\text{tan}^{-1}
  \frac{r_0}{\sqrt{\beta}}+\frac{r_0}{r_0^2
  +\beta}\right)+r_0;
\end{multline}
observe that $b(r_0)=r_0$, as required.

In this section, we would like to consider
microscopic wormholes with electric charge
$Q$.  Following Kim and Lee \cite{KL01},
we take the line element to be
\begin{equation}\label{E:line4}
ds^{2}=-e^{\nu(r)}dt^{2}+\frac{dr^2}
{1-\frac{b(r)}{r}+\frac{Q^2}{r^2}}
+r^{2}(d\theta^{2}+\text{sin}^{2}\theta\,
d\phi^{2}).
\end{equation}
where $b(r)$ is given by Eq.
(\ref{E:shape}).
In line element (\ref{E:line4}), the
effective shape function
$b_{\text{eff}}(r)$ is
\begin{equation}
   b_{\text{eff}}(r)=b(r)
        -\frac{Q^2}{r},
\end{equation}
where $b_{\text{eff}}(r_1)=r_1$ and
$r_1$ is the solution of the equation
$b(r)=r+Q^2/r$.  From Eq. (\ref{E:EQ2}),
we now get
\begin{equation}
   e^{\nu(r)}=\frac{1}{K}\left(\int^r_{r_1}
   \left(\frac{r'}{b_{\text{eff}}(r')}-1
   \right)^{-1/2}dr'\right)^2.
\end{equation}
Here we return to the discussion in Sec.
\ref{S:complete}.  Because of the event
horizon at $r=r_1$, the wormhole is not
traversable, thereby constituting an
Einstein-Rosen bridge.

Ref. \cite{LOR14} discusses microscopic
charged wormholes in the context of
quadratic Palatini gravity.  According
to this theory, the solution can be
interpreted as an electric flux going
through one mouth of the wormhole and
coming out of the other mouth.  The
result is a negative charge on one side
and a positive charge on the other.
Referring to Ref. \cite{MW57}, for all
practical purposes, there is no
difference between the kind of charge
described as a wormhole (i.e., by means
of a nontrivial topology) and a standard
point-like charge, suggesting that
spacetime could have a foam-like
structure.

For a detailed discussion of the
microstructure in conjunction with
entanglement and the $ER=EPR$ conjecture,
see Ref. \cite{LOR14}.
%END OF SECTION

\section{Summary}
An $n$-dimensional Riemannian space is
said to be of embedding class $m$ if
$n+m$ is the lowest dimension of the
flat space in which the given space can
be embedded.  A spherically symmetric
spacetime of embedding class two can be
reduced to class one by a suitable
transformation of coordinates.  From
the resulting metric (\ref{E:line3}),
we have
\[
  e^{\lambda}=1+\frac{1}{4}K
  e^{\nu}(\nu')^2,\quad K>0,
\]
where $K$ is a free parameter.  If we
start with a well-defined shape
function $b=b(r)$, then the resulting
wormhole has an event horizon and is
therefore not traversable.  Replacing
$b(r)$ by $m(r)$, the effective mass
inside a spherical star of radius $r$
and with $m(0)=0$, we obtain a valid
expression for $e^{\nu(r)}$ [Eq.
(\ref{E:EQ4})], thereby avoiding an
event horizon.  If $R$ is the radius
of the star, then the free parameter
$K$ allows a junction to an external
Schwarzschild spacetime at $r=R$.  The
potential $\nu(r)$ reduces to the
Newtonian limit $\nu=-m/r$ for large
$r$.  It is therefore not surprising
that spacetimes of embedding class one
have been used successfully for modeling
compact stellar objects.  As we have
seen, however, such spacetimes cannot
be used to model traversable wormholes
without introducing some additional
conditions.

Making use of a noncommutative-geometry
background, we can consider microscopic
wormholes with electric charge.  The
presence of an event horizon results
in an Einstein-Rosen bridge and is
therefore a viable model for a charged
particle.  The reason is that,
according to Ref. \cite{MW57}, for all
practical purposes, there is no
difference between a point-like charge
and a wormhole structure arising from
a nontrivial toplogy.  Ref. \cite{LOR14}
discusses the microstructure in
conjunction with entanglement and the
$ER=EPR$ conjecture.


\begin{thebibliography}{20}
\bibitem{MT88}M. S. Morris and  K. S. Thorne, Wormholes
   in spacetime and their use for interstellar travel:
   A tool for teaching general relativity, American
   Journal of Physics 56 (1988), 395-412.
\bibitem{MG17}S. K. Maurya and M. Govender, Generating physically
   realizable stellar structures via embedding, European
   Physical Journal C 77 (2017), ID: 347.     \
\bibitem{MM17}S. K. Maurya and S. D. Maharaj, Anisotropic
   fluid spheres of embedding class one using Karmarkar
   condition, European Physical Journal C 77 (2017), ID: 328.
\bibitem{MRG17}S. K. Maurya, B. S. Ratanpal and M. Govender,
   Anisotropic stars for spherically symmetric spacetime
   satisfying Karmarkar condition, Annals of Physics
   382 (2017), 36-49.
\bibitem{sM17}S. K. Maurya, Y. K. Gupta, S. Ray and D. Deb,
   A new model for spherically symmetric charged compact
   stars of embedding class 1, European Physical Journal
   C 77 (2017), ID: 45.
\bibitem{sM16}S. K. Maurya, Y. K. Gupta, S. Ray and D. Deb,
   Generalized model for anisotropic compact stars, European
   Physical Journal C 76 (2016), ID: 693.
\bibitem{sM19}S. K. Maurya, D. Deb and P. K. F. Kuhfittig
   A study of anisotropic compact stars based on embedding
   class 1 condition, International Journal of Modern
   Physics D 28 (2019), ID: 1950116.,
\bibitem {kK48}K. R. Karmarkar, Gravitational metrics of
   spherical symmetry and class one, Indian Academy of
   Sciences 27 (1948), 56-60.
\bibitem{SF20}M. Farasat Shamir and I. Fayyaz, Traversable
   wormhole solutions in $f(R)$ gravity via Karmarkar
   condition, European Physical Journal C 80 (2020),
   ID: 1102.
\bibitem{sS05}S. V. Sushkov, Wormholes supported by a
   phantom energy, Physical Review D 71 (2005), ID: 043520.
\bibitem{pK18}P. K. F. Kuhfittig, Two diverse models of
   embedding class one, Annals of Physics 392 (2018),
   63-70.
\bibitem{pK19}P. K. F. Kuhfittig, Spherically symmetric
   wormholes of embedding class one, Pramana - Journal
   of Physics 75 (2019), ID: 75.
\bibitem{MTW}C. W. Misner, K. S. Thorne and J. A. Wheeler,
   Gravitation, W. Freeman and Company, New York, 1973,
   p. 608.
\bibitem{GL09}R. Garattini and F. S. N. Lobo, Self-sustained
   traversable wormholes in noncommutative geometry,
   Physics Letters B 671 (2009), 146-152.
\bibitem{eW96}E. Witten, Bound states of strings and
   $p$-branes, Nuclear Physics B 460 (1996), 335-350.
\bibitem{SW99}N. Seiberg and E. Witten, String theory and
   noncommutative geometry, Journal of High Energy Physics
   9909 (1999), ID: 032.
\bibitem{SS03}A. Smailagic and E. Spallucci, Feynman
   path integral on the non-commutative plane,  Journal
   of Physics A 36 (2003), L-467-L-471.
\bibitem{NSS06}P. Nicolini, A. Smailagic and E. Spallucci,
   Noncommutative geometry inspired Schwarzschild black hole,
   Physics Letters B 632 (2006), 547-551.
\bibitem{NS10}P. Nicolini and E. Spallucci, Noncommutative
   geometry-inspired dirty black holes, Classical and
   Quantum Gravity 27 (2010), ID: 015010.
\bibitem{NM08}N. Nozari and S. H. Mehdipour, Hawking
   radiation as quantum tunneling from a noncommutative
   Schwarzschild black hole, Classical and Quantum
   Gravity 25 (2008), ID: 175015.
\bibitem{LL12}J. Liang and B. Liu, Thermodynamics of
   noncommutative geometry inspired BTZ black holes
   based on Lorentzian smeared mass distribution,
   Europhysics Letters 100 (2012), ID: 30001.
\bibitem{KL01}S.-W. Kim and H. Lee, Exact solutions
  of charged wormholes, Physical Review D 63 (2001),
  ID: 064014,
\bibitem{LOR14}F. S. N. Lobo, G. J. Olmo and D.
   Rubierra-Garcia, Microscopic wormholes and the
   geometry of entanglement, European Physical Journal
   C 74 (2014), ID: 2924.
\bibitem{MW57}C. W. Misner and J. A. Wheeler, Classical
   physics as geometry, Annals of Physics 2 (1957),
   525-603.

\end{thebibliography}
\end{document}